\documentclass[twocolumn,lettersize,journal]{IEEEtran}
\usepackage[utf8]{inputenc}
\usepackage{xcolor}
\usepackage{units}
\usepackage{mathtools}
\usepackage{amsmath}
\usepackage{amssymb}
\usepackage{graphicx}
\usepackage[unicode=true,
 bookmarks=false,
 breaklinks=false,pdfborder={0 0 1},backref=section,colorlinks=true]
 {hyperref}
\hypersetup{
 citecolor=blue,urlcolor=blue}

\makeatletter
\let\oldforeign@language\foreign@language
\DeclareRobustCommand{\foreign@language}[1]{%
  \lowercase{\oldforeign@language{#1}}}

\usepackage{amsfonts}\usepackage{algorithmic}
\usepackage{algorithm}
\usepackage{array}
\usepackage{textcomp}
\usepackage{stfloats}
\usepackage{url}
\usepackage{verbatim}
\usepackage{cite}
\usepackage{gensymb}\DeclarePairedDelimiter\bra{\langle}{\rvert}
\DeclarePairedDelimiter\ket{\lvert}{\rangle}
\DeclarePairedDelimiterX\braket[2]{\langle}{\rangle}{#1 \delimsize\vert #2}
\newcommand{\braketmatrix}[3]{\left \langle #1 \middle| #2 \middle| #3 \right \rangle}
\newcommand{\kket}[1]{\left| #1 \right\rangle\rangle}
\newcommand{\bbra}[1]{ \left\langle\langle #1 \right|}
\DeclareMathOperator{\Tr}{Tr}

\makeatother

\begin{document}
\global\long\def\br#1{\bra{#1}}%
\global\long\def\ke#1{\ket{#1}}%
\global\long\def\brke#1#2#3{\braketmatrix{#1}{#2}{#3}}%

\title{Discrete-Modulation Continuous-Variable Quantum Key Distribution with Probabilistic Amplitude Shaping over a Linear Quantum Channel}
\author{Emanuele Parente, Michele N. Notarnicola, Stefano Olivares, Enrico Forestieri,
\IEEEmembership{Life Member, IEEE}, Luca Potì, \IEEEmembership{Senior Member, IEEE}
and Marco Secondini, \IEEEmembership{Senior Member, IEEE} \thanks{Emanuele~Parente is with
the Telecommunications, Computer Engineering, and Photonics (TeCIP)
Institute, Scuola Superiore Sant'Anna, Pisa, Italy.}
\thanks{Enrico~Forestieri and Marco~Secondini are with
the Telecommunications, Computer Engineering, and Photonics (TeCIP)
Institute, Scuola Superiore Sant'Anna, Pisa, Italy, and also with
the National Laboratory of Photonic Networks, CNIT, Pisa, Italy.}
\thanks{Michele N. Notarnicola is with Department of Optics, Palacký University, 779 00 Olomouc, Czech Republic.}
\thanks{Stefano~Olivares is with Dipartimento di
Fisica “Aldo Pontremoli”, Università degli Studi di Milano, I-20133
Milano, Italy, and also with INFN, Sezione di Milano, I-20133 Milano,
Italy.}\thanks{Luca~Potì is with the National Laboratory of Photonic Networks, CNIT,
Pisa, Italy.}\thanks{Email: emanuele.parente@santannapisa.it.}}
\maketitle
\begin{abstract}
The practical implementation difficulties arising from the Gaussian modulation of the GG02 protocol lead us to investigate the possibilities offered by the combination of probabilistic amplitude shaping technique and quadrature amplitude modulation formats in the context of continuous variable quantum key distribution systems. Our interest comes from the fact that quadrature amplitude modulation and probabilistic shaping can be implemented with current technologies and are widely used in classical telecom equipment. In this treatment, we assume to work in the scenario of a linear quantum channel and we analyze maximum achievable secure key rates, maximum reachable distances and the resilience to noise of our discrete-modulation based protocol with respect to GG02, which is taken as a benchmark. In particular, we deal with the infinite key size regime, consider a homodyne detection scheme, and analyze what happens for different cardinalities of the input alphabet at different distances, in the case of collective attacks and in the reverse reconciliation picture. We find that our protocol, beyond being easily reproducible in the laboratory, provides a way to closely approach the theoretical performance offered by GG02 and, at the same time, preserves the ability to assure an unconditional security level.
\end{abstract}


\section{Introduction}

\IEEEPARstart{Q}{uantum} key distribution (QKD) is the most prominent application of quantum cryptography that ensures
the generation of a one time pad secure key between two authenticated remote parties, Alice and Bob, despite the presence of a powerful
adversary, Eve \cite{weedbrook2012gaussian}. The importance of QKD over classical cryptography lies in its ability to offer unconditional security, where no assumptions are made about the power of the adversary and whose strength is based only on the laws of quantum mechanics \cite{maurer1999information}.

QKD protocols come in the equivalent entanglement-based (EB) or prepare-and-measure (PM) versions: in both cases a quantum channel along which the exchange of optical signals occurs and an authenticated classical channel for the reconciliation stage are required. Eve has the power to tamper the communication over the quantum channel and to only listen to the conversation over the classical one \cite{scarani2009security}. In particular, the latter needs to be authenticated so that the legitimate parties are sure they are actually talking to each other; this can be achieved by means of information theoretically secure classical algorithms, so that the overall security of the protocol is not reduced \cite{martin2017introduction}. The classical channel is considered error free, so that all communication errors found during the post-processing stage are uniquely attributed to Eve’s interference along the quantum channel. By exploiting quantum features such as non-orthogonality or entanglement, in fact, the exchanged quantum states cannot be discriminated without disturbing the system, nor can they be perfectly copied as a consequence of the no-cloning theorem \cite{cariolaro2015quantum}, \cite{braunstein2005quantum}. In this way, Alice and Bob are not only able to detect the presence of an attacker, but are even able to estimate the amount of information an eavesdropper has acquired about the key, so that, whenever Eve’s interference exceeds a certain threshold, the protocol can be aborted. Since only the key is transmitted during the entire process, no information leakage occurs in this circumstance \cite{gisin2002quantum}, \cite{duvsek2006quantum}.
  
Due to its low interaction with matter and its longer decoherence time, light is the most common and practical choice to transmit information in QKD realm. The distinction of two families of QKD protocols, i.e. discrete-variable (DV-QKD) and continuous-variable (CV-QKD) ones, reflects the possibility to encode information about the key into different degrees of freedom of light: the polarization (or time bin) of single photons (or weak coherent states) in the former case and the orthogonal quadratures of an optical field in the latter. Subsequently to the signals exchange, Alice and Bob get a raw key that they refine after several post-processing steps, during which they exchange classical side information to eventually agree on a secure common key \cite{pirandola2020advances}.

Different aspects concur to assess the performance of a generic QKD protocol, i.e. the achievable distances, the rate at which the secure key is generated, the feasibility of its realization---including the costs of the employed devices---and the security level it provides \cite{wolf2021quantum}. Among the two classes mentioned above, CV-QKD protocols have shown a rapid growth nowadays \cite{zhang2024continuous}, primarily because they are able to circumvent the technological challenges of single photon sources and detectors required for DV-QKD \cite{kaur2021asymptotic}. Specifically, CV-QKD protocols are well-suited for use with fiber-optic systems (showing high performance in metropolitan areas) and exploit coherent detection, namely homodyne or heterodyne (also referred to as phase diversity or double homodyne) detection, which offers the advantages of higher detection efficiency, higher secret key rate (SKR) at short distances, low implementation costs, and greater compatibility with state-of-the-art telecom equipment (they are easily integrable with photonic-chips). Moreover, techniques such as the wavelength division multiplexing can be implemented at room temperature using standard hardware \cite{laudenbach2018continuous}, \cite{abushgra2022variations}, \cite{lupo2022quantum}, \cite{leverrier2009unconditional}.

The most famous protocol in this category and which is currently used as a benchmark in CV-QKD is GG02, introduced by Grosshans and Grangier in 2002 \cite{grosshans2002continuous}, and exploiting coherent states with random amplitude drawn from a Gaussian distribution that are sent through a noisy untrusted channel. The choice of Gaussian modulation (GM) is suggested as it achieves the Shannon capacity of a Gaussian channel between Alice and Bob, in the presence of coherent detection at the receiver \cite{cariolaro2015quantum}, \cite{holevo2012quantum}. Although the use of GM simplifies security proofs \cite{weedbrook2012gaussian}, \cite{laudenbach2018continuous}, \cite{cerf2007quantum}, \cite{notarnicola2024quantum} it entails several practical difficulties: handling a continuous modulation requires an infinite resolution of the digital-to-analog converter (DAC) at the transmitter and of the analog-to-digital converter (ADC) at the receiver, in addition to the requirement of an infinite extinction ratio of the modulator and to an extreme computational burden for the random number source on the transmitter side. Furthermore, signal modulation with infinite peak power is theoretically required to generate a Gaussian distribution, which indeed has infinite support in phase-space \cite{kaur2021asymptotic}, \cite{jouguet2012analysis}, \cite{diamanti2015distributing}. Then, the resulting lack of an efficient error-correction procedure at both high and low signal-to-noise ratio (SNR) limits range and feasibility of this protocol \cite{leverrier2009unconditional}.

A practical alternative to GM is offered by discrete modulation formats, which are commonly used in digital communications and can be implemented using simple modulation schemes and readily available optical components  \cite{djordjevic2019discretized}. An interesting choice is quadrature amplitude modulation (QAM), whose constellation points in phase space are arranged on a square lattice with finite support. This structure enables practical implementation using a finite-power laser and a nested Mach--Zehnder modulator \cite{proakis2001digital}. QAM is widely employed in optical communications, as it offers high spectral efficiency with reasonable implementation complexity \cite{winzer2012high}. Its performance can be further enhanced by combining it with probabilistic amplitude shaping (PAS), where transmitted symbols are selected from the constellation with probabilities that depend on their energy, with the goal of maximizing the bit rate for a given mean energy per symbol \cite{kschischang1993optimal}. Key advantages of PAS-QAM include its fine rate granularity and its ability to operate near Shannon capacity over the additive white Gaussian noise (AWGN) channel across a wide range of SNRs, using only standard binary error correction codes and bit-wise soft decoders within a pragmatic bit-interleaved coded modulation (BICM) framework \cite{bocherer2015bandwidth}. Owing to these properties and its simplicity, PAS-QAM has become a widely adopted modulation scheme for high-capacity fiber-optic communication systems \cite{buchali2015rate}, \cite{fehenberger2018information} \cite{fehenberger2016probabilistic}. 

Incorporating the advantages of PAS-QAM into a practical CV-QKD setup would represent a significant step toward the large-scale development of QKD systems based on mature technologies already deployed in current optical networks. For instance, the feasibility of using binary codes and bit-wise soft decoders to implement an efficient reverse reconciliation protocol for CV-QKD with QAM modulation was recently demonstrated in \cite{origlia2025soft}. However, such a change in the protocol should be supported by an accurate security analysis. The challenging part in establishing the security of CV-QKD protocols comes from the fact that an infinite dimensional Hilbert space is involved and this prevents the possibility of a complete tomography; therefore, a precise knowledge of the state Alice and Bob are actually sharing is impossible \cite{weedbrook2012gaussian}. In fact, even though the individual pulses are Gaussian states, the average quantum state is non-Gaussian, so that the covariance matrix estimation performed during the channel evaluation stage is not sufficient to fully characterize it. However, in 2006 Navascues \textit{et al.} \cite{navascues2006optimality} derived a lower bound on the SKR obtained by considering the Gaussian state associated with the same covariance matrix of the actual non-Gaussian state shared by Alice and Bob, providing a well-suited security framework for very large QAM constellations. Tighter bounds, also required for the investigation of a more realistic scenario of non-asymptotic key length exchange, can be found using convex optimization problems and more precisely semidefinite programming (SDP); on cons, these methods are numerically intensive and unsuitable for the large constellations size cases \cite{ghorai2019asymptotic}, \cite{lin2019asymptotic}. Recently, Denys \textit{et al.} \cite{denys2021explicit} addressed this problem and introduced an explicit numerical bound for the asymptotic secret key rate computation of a coherent-state CV-QKD protocol over an arbitrary quantum channel: their approach exploits a combination of Gaussian-optimality arguments and semidefinite programming, and their results yield an explicit bound that is tight for QAM constellations with 64 or more symbols.

In this work, we propose a QAM based CV-QKD protocol implemented over a linear quantum channel, that well models transmission through common optical fibers, and study the achievable rates, achievable distances and maximum tolerable noise in the case of collective attacks, reverse reconciliation scheme, infinite key size, homodyne detection, and in the distance range $\unit[0.5\text{--}300]{km}$. We consider different constellation sizes and compare the above-mentioned figures of merit in the cases where the input symbols are extracted according to a uniform distribution (U-) or a non-uniform distribution through the probabilistic amplitude shaping technique (PAS-), with those corresponding to GG02. We then compare the SKR obtained under this unconditional security approach with the wiretap channel assumption outlined in \cite{notarnicola2023probabilistic}, where Eve is assumed to implement a beam-splitting attack, and highlight the advantages, in terms of performance, adduced by the use of the PAS technique in a CV-QKD protocol with QAM modulation and coherent states.

\section{Methods}

\subsection{Proposed protocol}\label{subsec:Proposed_protocol}

\begin{figure}[htp]
    \centering
    \includegraphics[width=1\columnwidth]{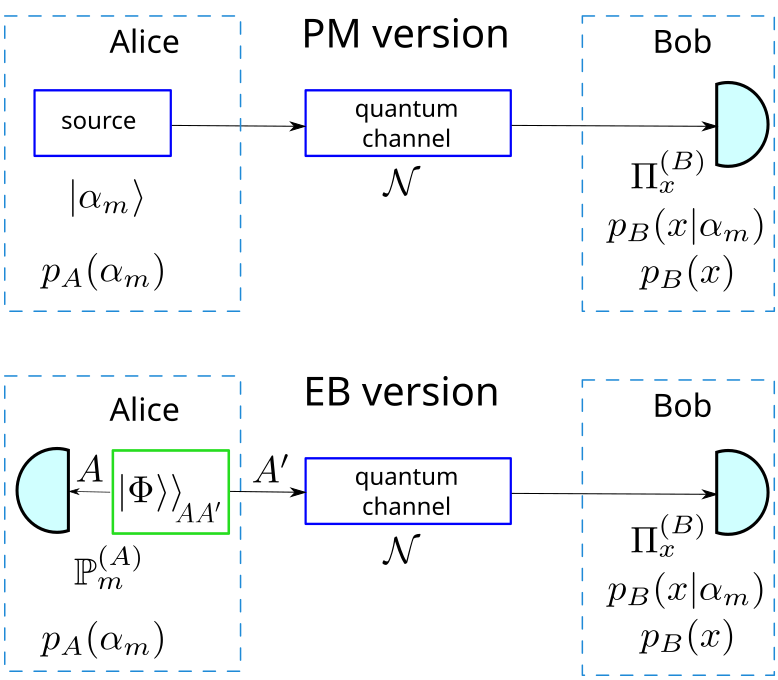}
    \caption{Prepare-and-measure version (top) and entanglement-based version (bottom) of our discrete modulated protocol.}
    \label{fig:PM_EB}
\end{figure}

\noindent With reference to Fig. (\ref{fig:PM_EB}), we describe now the PM version of our discrete modulated (DM) protocol and consider the asymptotic limit in which an infinite number of symbols is transmitted. 

Alice (A) encodes a pair of classical variables $(x_A,y_A)$ in a set of displaced single-mode coherent states $\ket{x_A + i y_A}$, where $x_A, y_A \in \mathbb{R}$ are generated according to a discrete-valued probability distribution. Then, Alice sends these states to Bob (B) through a quantum channel under control of Eve (E). On the receiver side, Bob performs a quantum homodyne detection, measuring at random one of the two orthogonal quadratures $q$ or $p$, which satisfy the commutation relation $[q,p] = 2 i \sigma_0^2$, where $\sigma_0^2$ is the shot-noise variance. In the following, we adopt shot-noise units (SNU), that is, we set $\sigma_0^2 = 1$; moreover, without loss of generality, we assume that Bob measures the quadrature $q$ through the projector $\Pi_x^B = \ket{x}\bra{x}$.

In this work, we are interested in the transmission of signals via a fiber optic link; this medium is characterized by the transmittance parameter $T$, which reflects the amount of losses caused by random scattering of the light passing through it and which depends exponentially on the length $d$ of the communication link according to $T=10^{-\eta d/10}$, where $\eta=\unit[0.2]{dB/km}$ is the attenuation coefficient at the optical wavelength of $\unit[1550]{nm}$ \cite{scarani2009security}. Beyond losses, also the noise affects the reliability of communication, and its impact is quantified by the excess noise parameter $\xi$. In our analysis, we assume that the quantum signal transmission and detection may be affected by the presence of noise, and that the quantum channel is linear. A linear quantum channel represents a class of channels that includes also the category of Gaussian channels \cite{weedbrook2012gaussian}, \cite{cariolaro2015quantum}. In fact, this channel is characterized by the input-output relations for the quadrature operators in the Heisenberg picture \cite{zhang2024continuous}, \cite{notarnicola2024quantum}

\begin{equation}{\label{lin_transfor}}
    {q}_{B} = \sqrt{T} \hspace{0.1cm} {q}_{A'} + {q}_b
    \hspace{0.4cm}\textnormal{and}\hspace{0.4cm}
    {p}_{B} =\sqrt{T} \hspace{0.1cm} {p}_{A'} + {p}_b,
\end{equation}

\noindent where $q_b$ and $p_b$ represent the quadrature operators of some additional noise modes, assumed to be uncorrelated with the input quadratures $q_{A'}$ and $p_{A'}$ (the input signal) and being characterized by the first two order moments

\begin{equation}
    \begin{cases}
        \langle {q}_b \rangle = \langle {p}_b \rangle = 0 \hspace{0.1cm},\\
        \langle {q}_b^2 \rangle = \langle {p}_b^2 \rangle = 1-T+\xi T \hspace{0.1cm}.
    \end{cases}
\end{equation}

After the signal exchange, Alice and Bob perform a post-processing stage in which they share some side information through a classical authenticated channel to agree on a secret common key. Specifically, they first perform a parameter estimation operation, where they estimate both the experimental parameters $T$ and $\xi$ describing the channel and the amount of information Eve may have acquired so far; then, a reconciliation step for error correction---which can be either direct (DR) or reverse (RR), depending on whether the information on the key bits is shared by Alice or Bob respectively---and, finally, a privacy amplification process where they extract from the raw key a shorter secret key, eliminating any possible information gained by the eavesdropper \cite{pirandola2020advances}, \cite{wolf2021quantum}. Here we adopt a RR scheme, which is more advantageous, since it overcomes the $\unit[3]{dB}$ limit problem of the DR scheme. \cite{martin2017introduction}, \cite{wolf2021quantum}, \cite{laudenbach2018continuous}.

Finally, to ensure a higher level of security to our protocol, we consider the untrusted-device scenario---where Eve has access to both channel and Bob's apparatus---and the case of collective attacks, which are the most powerful in the asymptotic limit of an infinitely long key. In the collective attack picture, Eve performs an independent and identically distributed (i.i.d.) attack on all signals by making these interact with a set of separable ancilla states, stores each of these ancillae in a quantum memory and jointly measures them at the end of the post-processing. In this way, Eve can take advantage of both signal transmissions and the public information Alice and Bob have shared along the classical channel during post-processing.

\begin{figure}[htp]
    \centering
    \includegraphics[width=1\columnwidth]{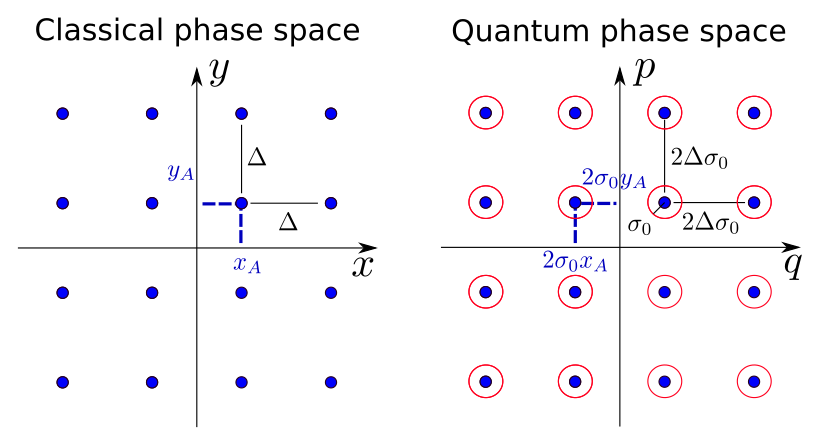}
    \caption{16QAM constellation in classical (left) and quantum phase space (right): the symbols are spaced by $\Delta$ and centered at $(x_A,y_A)$ in the former case, spaced by $2 \Delta$ and placed in $(2 \sigma_0 x_A, 2 \sigma_0 y_A)$ in the latter.}
    \label{fig:16QAM}
\end{figure}

\subsection{Alice's modulation}\label{subsec:Alice_modulation}

During the modulation stage, Alice generates couples of quadrature values $(x_A, y_A)$ representing the coordinates of a uniform square $M$-ary QAM constellation in classical phase space: these values belong to the finite alphabet $\mathcal{A} = \Lambda \times \Lambda$, where $\Lambda = \{ n \Delta \hspace{0.1cm} | \hspace{0.1cm} n = -(N-1)/2, .. , (N-1)/2 \}$ contains $N=2^{k}$ points, $k = \mathbb{N}$ and $\Delta \in \mathbb{R}$, the latter being called the scaling factor or the symbol spacing. Thereafter, Alice maps the $M = N^2$ constellation points of $\mathcal{A}$ in quantum phase space \cite{olivares2012quantum} by encoding each couple $(x_A, y_A)$ into the corresponding coherent state $\ket{x_A + i y_A}$. In particular, the distance between neighbouring points in the classical phase space is equal to $\Delta$; in turn, as a consequence of the expectation values of the quadrature operators over a generic coherent state $\ket{\alpha}$ \cite{laudenbach2018continuous}
\begin{equation}
    \langle {q} \rangle_{\alpha} = 2 \sigma_0 Re (\alpha)
    \hspace{0.6cm}\textnormal{and}\hspace{0.6cm}
    \langle {p} \rangle_{\alpha} = 2 \sigma_0 Im (\alpha),
\end{equation} 
we find that, in quantum phase space, the QAM symbols are still arranged in a square lattice, but they appear centered at $(2 \sigma_0 x_A \hspace{0.1cm}, 2 \sigma_0 y_A)$ and spaced by $2 \sigma_0 \Delta$, see Fig. (\ref{fig:16QAM}).

The value of $\Delta$ is related to the average energy of the constellation, here expressed in terms of the average number of photons
\begin{equation}{\label{Ener}}
    \dfrac{\bar{n}}{2} = \sum_{z=-(M-1)/2}^{(M-1)/2} p_A(z) \hspace{0.1cm} |z|^{2},  
\end{equation}
\noindent where $p_{A}(z)$ is the input mass probability governing the extraction of the quadrature component $z = (x_A,y_A)$. One trivial choice for $p_{A}(z)$ is the uniform distribution; anyway, compared to other modulation formats that still require coherent detection such as phase shift keying (PSK), QAM-based systems can be enriched by the use of the PAS technique. The combination of QAM and PAS offers a way to improve the bit rate of the communication with fixed launch power. In particular, using the maximum entropy principle \cite{cover1999elements} or the variational calculus \cite{forney1989multidimensional} one can find that Maxwell-Boltzmann (MB) distribution is the one that maximizes the Shannon entropy for a fixed average energy of a discrete constellation (and hence for a certain power usage), being able to provide the ultimate shaping gain of $\unit[1.53]{dB}$ \cite{forney1984efficient}. The MB distribution can be written as
\begin{equation}{\label{MB}}
    \mathcal{M}_{\beta}(z) = \dfrac{e^{-\beta z^2}}{\sum_{z} e^{-\beta z^2}} \hspace{0.3cm},
\end{equation}
\noindent where $\beta$ is a shaping parameter: when $\beta\rightarrow0$ we get a uniform distribution, while for $\beta\gg1$, a general $M$-QAM reduces to a QPSK format, since the extraction of the outer points of the constellation is ruled out \cite{kschischang1993optimal}.

In this work we compare the performance of our protocol considering both uniform and non uniform signaling (that is, PAS using the MB distribution) of quantum states. In the uniform case, i.e. $p_A(z) = 1/N$, the constellation is characterized by a single parameter, the spacing $\Delta$ between the symbols, which is completely specified by the choice of the average number of photons $\bar{n}$: given the latter, $\Delta$ can be easily found by inverting (\ref{Ener}). Instead, when the a priori probabilities follow the MB distribution (\ref{MB}), the modulation is characterized by two parameters, the spacing  $\Delta$ and the shaping parameter $\beta$. These two degrees of freedom are constrained by the desired average number of photons $\bar{n}$: for any fixed $\beta$, the corresponding value of $\Delta$ can be obtained via (\ref{Ener}). The optimal value of $\beta$ is then determined by maximizing the SKR (see next section), according to the numerical procedure described in \cite{bocherer2015bandwidth}.

\subsection{Secret key rate calculation\label{subsec:Secrete-key-rate-calculation}}

The asymptotic SKR is a fundamental metric to assess the performance and security of a CV-QKD protocol. Although the assumption of an infinitely long signal exchange is not  realistic, it provides an upper bound on the achievable SKR and simplifies the theoretical security analysis \cite{diamanti2015distributing}. 

For collective attacks with the RR scheme, the SKR is bounded from below by the Devetak-Winter formula \cite{pirandola2020advances}

\begin{equation}
    \mathit{SKR} \geq \zeta I_{AB}-\chi_{BE}, {\label{DW}}
\end{equation}

\noindent where $I_{AB}$ is the mutual information shared by Alice and Bob, $\chi_{BE}$ is the Holevo information (HI) between Bob and Eve, while $\zeta \in [0,1]$ is the reconciliation efficiency. In particular, this parameter takes into account an imperfect error correction performed by Alice and Bob and then limits the classical amount of information that they can extract; on the other hand, the HI term is not affected by $\zeta$ since Eve's power is assumed not to be limited by current technology \cite{leverrier2009unconditional}. From now on, we set $\zeta$ to the realistic value of $0.95$ \cite{denys2021explicit}. \\

\subsubsection{Mutual information}

the mutual information $I_{AB}$ measures the average information content exchanged between Alice and Bob \cite{maurer1999information}. In order to compute this term, we consider the PM scheme described in \ref{subsec:Proposed_protocol}. For each use of the channel, Alice prepares the coherent state $\ket{x_A + i y_A}$ and sends it through the quantum channel; at the output, Bob measures the quadrature $q$ and obtains the outcome $x_B$. Given the linearity of the channel, equivalent to an AWGN picture, the random variables are described by the general input-output relation $x_B= 2\sigma_0 \sqrt{T}x_A+n$, where the noise $n$ is normally distributed $\mathcal{N}(0,\sigma_{\xi}^{2})$; in our case, $x_A$, $n$ and hence $x_B$ are real numbers. For a generic $M$-ary modulation alphabet, the mutual information can then be expressed as \cite{secondini2020information} 

\begin{equation*}
    I_{AB} = \sum_{x_A}\int dx_B \hspace{0.1cm}p_A(x_A)p_{B|A}(x_B|x_A) \times
\end{equation*}
\begin{equation}
    \log_{2}\dfrac{p_{B|A}(x_B|x_{A})}{\sum_{x_A'} p_A(x'_{A})p_{B|A}(x_B|x'_{A})},
\end{equation}

\noindent where $p_A(x_A)$ is the \emph{a priori} probability of the classical symbol $x_A$ and $p(x_B|x_A)$ is the transition probability relating the input and the output of the AWGN channel  
\begin{equation}{\label{trans_prob}}
    p_{B|A}(x_B|x_A)=\dfrac{1}{\sqrt{2\pi\sigma_{\xi}^{2}}}\exp\left(-\dfrac{(x_B- 2 \sigma_0 \sqrt{T} x_A)^{2}}{2\sigma_{\xi}^{2}}\right),
\end{equation}
\noindent where the value of the variance $\sigma_{\xi}^{2} = \sigma_0^2(1 + T \xi)$ in (\ref{trans_prob}) quantifies the impact of noise on the signal received at the channel output.\\

\subsubsection{Holevo information}

\noindent the Holevo information term $\chi_{BE}$ quantifies the maximum amount of information about Bob's measurement outcomes that Eve can obtain in the reverse reconciliation scenario \cite{zhang2024continuous}. An accurate determination of this quantity is essential for the security analysis of our CV-QKD protocol. 

To facilitate the analysis, we exploit the equivalence between the PM and EB descriptions of QKD protocols \cite{laudenbach2018continuous},  and focus on the latter, which offers a more tractable theoretical framework. If in the PM scheme Alice prepares the coherent states $\{ \ket{\alpha_k} , p_A(\alpha_k) \}$, where $k = (x_A,y_A) \in \mathcal{A}$ and $\mathcal{A}$ the alphabet of our QAM constellation, the latter described by the density operator $\rho = \sum_k p_A(\alpha_k) \ket{\alpha_k}\bra{\alpha_k}$, in the EB scheme Alice actually produces a set of a bipartite entangled states $\kket{\Phi}_{AA'}$ in the two modes $A$ and $A'$ \cite{notarnicola2024quantum}, \cite{lin2019asymptotic}
\begin{equation}{\label{bipartite_state}}
    \kket{\Phi}_{AA'} = \sum_{k \in \mathcal{A}} \sqrt{p_A (\alpha_k)} \hspace{0.1cm} \ket{\varphi_k}_A \ket{\alpha_k}_{A'},
\end{equation}
where $\{\ket{\varphi_k}_A\}_{k \in \mathcal{A}}$ form an orthonormal basis in mode $A$ on which Alice, in order to determine which state to send to Bob, performs a local measurement described by POVM $\{\Pi_k^A = \ket{\varphi_k}_A\bra{\varphi_k} \}_{k \in \mathcal{A}}$. This causes the state $\ket{\alpha_k}_{A'}$ on the other mode $A'$ to be projected onto a coherent state with the corresponding probability $p_A(\alpha_k)$.

Thereafter, she sends the state $\ket{\alpha_k}_{A'}$ to Bob over the quantum channel, the latter described by a completely positive (CP) quantum map $\mathcal{N}_{A' \rightarrow B}$; the bipartite state ${\rho}_{AB}$ shared by Alice and Bob is then
\begin{equation}
    {\rho}_{AB} = (\mathbb{I}_A \otimes \mathcal{N}_{A' \rightarrow B}) \left[ \hspace{0.1 cm} \kket{\Phi}_{AA'}\hspace{-0.05 cm}\bbra{\Phi} \hspace{0.1 cm} \right],
\end{equation}
where $\mathbb{I}_A$ is the identity channel acting on the states collected at Alice's side and $(\mathbb{I}_A \otimes \mathcal{N}_{A' \rightarrow B})$ is a completely positive trace preserving (CPTP) map. The conditional state Bob receives after Alice's projective measurement is:
\begin{equation}
    {\rho}_{B|\alpha_k} = \dfrac{1}{p_A(\alpha_k)} \Tr_A \left[ {\rho}_{AB} \hspace{0.1cm} (\Pi_k^A \otimes \mathbb{I}_B) \right].
\end{equation}
If Alice has sent the state $\ket{\alpha_k} = \ket{x_A + i y_A}$, for Bob performing a homodyne detection and measuring the quadrature $q$ with POVM $\Pi_x^B = \ket{x}\bra{x}$, the conditional distribution of Bob's measurement outcome $x_B$ is given by
\begin{equation}
    p_{B|A}(x_B|x_A) = \Tr_A \left[ {\rho}_{AB} \hspace{0.15 cm} \ket{\varphi_k}\bra{\varphi_k}_A \otimes \Pi_x^{B} \right],
\end{equation}
and the corresponding marginal distribution of Bob's outcome is:
\begin{equation}
    p_B(x_B)=\sum_m p_{B|A}(x_B|x_A)p_A(x_A).
\end{equation}

Unfortunately, these available statistics---namely, $p_A(x_A)$, $p_{B|A}(x_B|x_A)$, and $p_B(x_B)$---are not sufficient for the proper parties to uniquely characterize the underlying untrusted quantum channel, and this lack of complete information plays in favor of Eve. In fact, the eavesdropper can exploit the non-uniqueness of the noisy maps $\mathcal{N}$ that yield the same input-output statistics by selecting and implementing the unitary dilation of the quantum channel that maximizes her accessible information while remaining consistent with the statistics observed by Alice and Bob, hence without being noticed. For the purpose of unconditional security, we consider Eve to perform a purification attack; then, the Devetak-Winter formula can be re-written as \cite{ghorai2019asymptotic}
\begin{equation}
    \mathrm{SKR}=\zeta I_{AB}-\sup_{\mathcal{U}_{A'\rightarrow BE\in S}}\chi_{BE}\label{SKR},
\end{equation}
where the supremum is taken over all isometric quantum channels controlled by Eve and belonging to the set $S$ compatible with the channel statistics obtained by Alice and Bob during the parameter estimation phase of the PM version of the protocol. In the case of the purification attack here considered, Eve holds a tripartite isolated system $ABE$ that is described by the pure state $\ket{\psi}_{ABE}$ and of which she controls the unitary dilation additional modes of the noisy CP map $\mathcal{N}$. The density matrices ${\rho}_{AB}$ and ${\rho}_{E}$ can be found from $\ket{\psi}_{ABE}$ by tracing out the modes $E$ and $AB$, respectively.

Now, the HI term corresponds to \cite{pirandola2020advances}
\begin{equation}
    \chi_{BE}= S(E) - S(E|B),
\end{equation}
where $S(E)$ is the von Neumann entropy of Eve's state $\rho_{E}$ accessible to Eve for collective measurements, and $S(E|B)$ is the von Neumann entropy conditional to the same state after Bob's measurement (a homodyne projection in our case). However, in a purification attack scenario the state $\rho_{AB}$ provides the same entropic calculations of $\rho_E$, i.e. $S(E) = S(AB)$ and $S(E|B) = S(A|B)$, so that
\begin{equation}{\label{Holevo_Information}}
    \chi_{BE}= S(E) - S(E|B) = S(AB) - S(A|B).
\end{equation}
This implies that $\chi_{BE}$ can be directly computed from the bipartite state $\rho_{AB}$ shared between Alice and Bob \cite{weedbrook2012gaussian}, \cite{pirandola2020advances}, \cite{laudenbach2018continuous}.

Given this consideration, we provide the security analysis by invoking the "optimality of Gaussian attacks" theorem \cite{notarnicola2024quantum}, \cite{navascues2006optimality}, \cite{garcia2006unconditional}, \cite{notarnicola2023long}, \cite{leverrier2009theoretical}, that bounds the HI in (\ref{Holevo_Information}) as $\chi_{BE} < \chi_{BE}^{(\rm G)}$, where $\chi_{BE}^{(\rm G)}$ is the HI associated with the Gaussian state $\rho_{AB}^{(\rm G)}$ having the same covariance matrix $\Gamma_{AB}$ of $\rho_{AB}$. Up to local operations, $\Gamma_{AB}$ is expressed as 
\begin{equation}{\label{cov_matrix_general}}
    \Gamma_{AB}=\begin{pmatrix}V\mathbb{I}_{2} & Z\sigma_{z}\\
    Z\sigma_{z} & W\mathbb{I}_{2}
    \end{pmatrix}=\begin{pmatrix}\gamma_{A} & \gamma_{C}\\
    \gamma_{C} & \gamma_{B}
    \end{pmatrix},
\end{equation}
where $\mathbb{I}_{2}=\mathop{\mathrm{diag}}(1,1)$ is the identity matrix and $\sigma_{z}=\mathop{\mathrm{diag}}(1,-1)$ the Pauli matrix, $V$ is the variance at Alice's side (of mode $A$) and $W$ the one at Bob's side, while $Z$ is the correlation term. 
At this point, the von Neumann entropy of the joint state $\rho_{AB}^{(\rm G)}$ can be obtained as\cite{ghorai2019asymptotic}, \cite{olivares2012quantum}
\begin{equation}
    S({\rho}_{AB})=g \left(\frac{d_1-1}{2} \right) + g \left(\frac{d_2-1}{2} \right),
\end{equation}
where $d_{1,2}$ are the symplectic eigenvalues of $\Gamma_{AB}$ satisfying the constraint $d_{1,2}\geq1$, namely:
\begin{equation}
    d_{1,2}=\sqrt{(\Delta\pm\sqrt{\Delta^{2}-4I_{4}})/2},
\end{equation}
and
\begin{equation}
    g(x)= (x+1) \log_2 (x+1) - x \log_2 x ,
\end{equation}
while $\Delta=I_{1}+I_{2}+2I_{3}$, and
\begin{equation*}
    I_{1}=\det(\gamma_{A})\hspace{0.15cm},\hspace{0.15cm}I_{2}=\det(\gamma_{B})\hspace{0.15cm},
\end{equation*}
\begin{equation}
    \hspace{0.15cm}I_{3}=\det(\gamma_{C})\hspace{0.15cm},\hspace{0.15cm}I_{4}=\det(\Gamma_{AB}).
\end{equation}

The conditional entropy $S(\rho_{A|B}^{(\rm G)})$, instead, can be computed as $S(\rho_{A|B}^{(\rm G)})=g(d_{3})$ \cite{notarnicola2024quantum} with
\begin{equation}
    d_{3}=\det(\gamma_{A}-\gamma_{C}(\gamma_{B}+\gamma_{M})^{-1}\gamma_{C}^{T}),
\end{equation}
and $\gamma_{M}$ is the $2 \times 2$ covariance matrix associated with homodyne detection \cite{ferraro2005gaussian}:
\begin{equation}
\gamma_{M}=\lim_{\lambda\rightarrow0}
    \begin{pmatrix}\lambda & 0\\
        0 & \frac{1}{\lambda}
    \end{pmatrix}.
\end{equation}

In practical implementations of CV-QKD, while Alice's variance $V=1+2\bar{n}$ is known by the modulation stage, Bob's variance $W$ is determined via measurements (the transmittance and excess noise values are accessible after the parameter estimation stage), so the only unknown is the cross term $Z$, that needs nonlocal bipartite measurements to be directly evaluated \cite{zhang2024continuous}.

In this regard, the hypothesis of a linear quantum channel solves the problem and provides an exact evaluation of $Z$ and then of the covariance matrix under a purification attack. The linearity of the channel ensures that the covariance matrix $\Gamma_{AB}$ of the state shared by Alice and Bob can be expressed as a function of the covariance matrix $\Gamma_{AA'}$ of the bipartite state $\kket{\Phi}_{AA'}$ generated by Alice, which reads
\begin{equation}
    \Gamma_{AA'} =
    \begin{pmatrix}
        V \mathbb{I}_2 & Z_{AA'} \hspace{0.1cm} \sigma_z\\
        Z_{AA'} \hspace{0.1cm} \sigma_z & V \mathbb{I}_2
    \end{pmatrix},
\end{equation}
in which
\begin{equation}
        Z_{AA'} = \dfrac{1}{2} \hspace{0.1cm} {}_{AA'}\bbra{\Phi} ({q}_A {q}_{A'} - {p}_A {p}_{A'}) \kket{\Phi}_{AA'} .
\end{equation}

In order to find $Z$, we use the Schmidt decomposition of the bipartite entangled state $\kket{\Phi}_{AA'}$ \cite{denys2021explicit}:
\begin{equation}
    \kket{\Phi}_{AA'} = \left( (\rho_A^{*})^{1/2} \otimes \mathbb{I} \right) \kket{EPR}_{AA'} {\label{PHI_2}}\hspace{0.1cm},
\end{equation}
with $\kket{EPR} = \sum_{n=0}^{\infty} \ket{n_A}\ket{n_{A'}}$ the two-mode squeezed vacuum state with infinite squeezing and $\{ \ket{n}_A \}_{n=0}^{\infty}$, $\{ \ket{n}_{A'} \}_{n=0}^{\infty}$ the Fock basis for the modes $A$ and $A'$. Straightforward calculations lead to:
\begin{equation*}
    Z_{AA'} =  
\end{equation*}
\begin{equation}
    \sum_{mn} \braketmatrix{mm}{ [{\rho}_A^{1/2} \otimes \mathbb{I}_2] ({a}_A {a}_{A'})  [({\rho}_A^{*})^{1/2} \otimes \mathbb{I}_2]}{nn} + h.c. =
\end{equation}
\begin{equation}
    = \sum_{mn} \braketmatrix{m}{ {\rho}_A^{1/2} {a}_A ({\rho}_A^{*})^{1/2} } {n} \braketmatrix{m}{{a}_{A'}}{n} + h.c. ,
\end{equation}
where $\braketmatrix{m}{{a}_{A'}}{n} = \sqrt{n} \hspace{0.1cm} \delta_{m, n-1}$ so that
\begin{equation}
    Z_{AA'} = \sqrt{n} \sum_{n} \braketmatrix{n-1}{ \hspace{0.1cm} {\rho}_A^{1/2} {a}_A ({\rho}_A^{*})^{1/2} } {n} + h.c. = 
\end{equation}
\begin{equation}
    = \braketmatrix{n}{ {a}_A^{\dagger} \hspace{0.1cm} {\rho}_A^{1/2} \hspace{0.1cm}  {a}_A \hspace{0.1cm}  ({\rho}_A^{*})^{1/2} } {n} + h.c.,
\end{equation}
and
\begin{equation}
    Z_{AA'} = 2\mathop{\mathrm{Tr}}[\rho_{A}^{1/2}\,{a}\hspace{0.05cm}\rho_{A}^{1/2}\,{a}^{\dagger}].
\end{equation} 
Eventually, the covariance matrix reads \cite{notarnicola2024quantum}
\begin{equation}
    \Gamma_{AB}=\begin{pmatrix}V\mathbb{I}_{2} & \sqrt{T} \hspace{0.1cm} Z_{AA'}\sigma_{z}\\
    \sqrt{T} \hspace{0.1cm} Z_{AA'} \sigma_{z} & TV_B\mathbb{I}_{2}
    \end{pmatrix},
\end{equation}
where
\begin{equation}
    V_B=V+\chi \hspace{0.5cm},\hspace{0.5cm}\chi=\dfrac{1-T}{T}+\xi.
\end{equation}

\section{Results}

\noindent The performance of the proposed  CV-QKD protocol, which employs QAM modulation combined with PAS, is evaluated by comparing it against two baselines: (i) QAM with uniform distribution, and (ii) the benchmark GG02 protocol, which uses GM. Specifically, we will analyze and compare the maximum achievable SKRs, the maximum transmission distances, and the maximum tolerable excess noise across these schemes. Moreover, for the same protocols, we will investigate the optimal launch power and the power range over which a positive SKR can be achieved, both expressed in terms of photons per symbol on Alice's side.  

\subsection{Maximum achievable SKR}

\begin{figure*}[htp]
    \centering
    \includegraphics[width=1\textwidth]{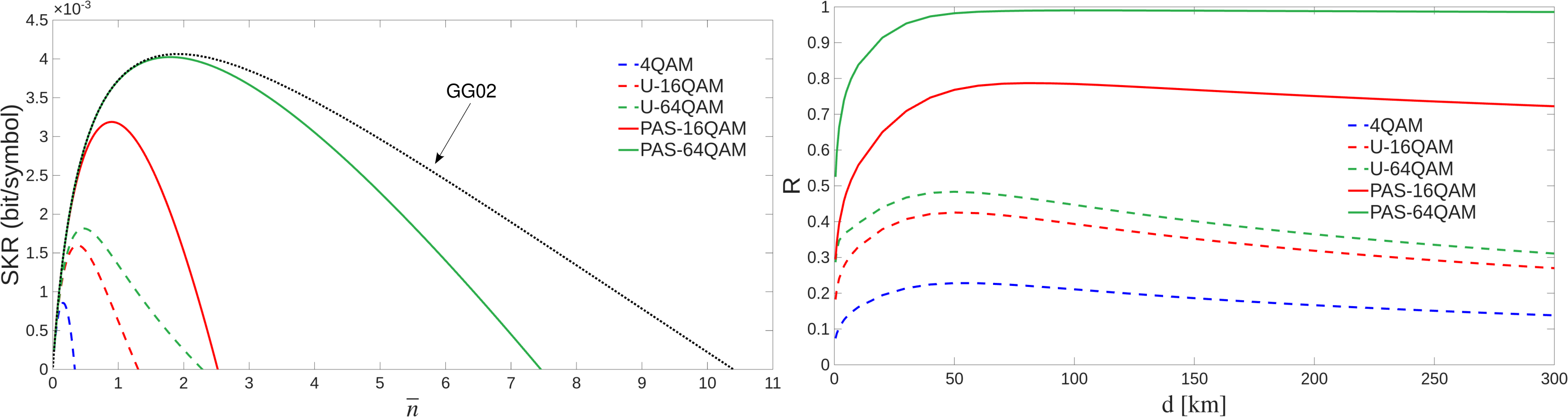}
    \caption{SKR for GG02 and for 4QAM, 16QAM, 64QAM with uniform and non-uniform input signaling in terms of the number of photons on Alice's side at $\unit[100]{km}$ in the case of collective attacks, homodyne detection, reverse reconciliation and no excess noise (left); corresponding values of $\mathrm{R}$ in the range $\unit[0.5\text{--}300]{km}$ (right).}
    \label{fig:100km}
\end{figure*}

\noindent We begin by addressing the noiseless scenario ($\xi=0$). Fig. (\ref{fig:100km}) (left) shows the SKR as a function of the average number of photons $\bar{n}$ at Alice's side for GG02 and for 4, 16 and 64QAM with uniform (U) distribution or PAS, at a fixed distance of $\unit[100]{km}$. As expected, different constellations yield different SKR performance. 4QAM, equivalent to QPSK, achieves the lowest SKR peak at a very low $\bar{n}$, after which the SKR quickly decays to zero. The SKR peak and the support of the SKR curve in the photon number domain can be increased by increasing the cardinality of the constellation. However, with uniform signaling the improvement is quite limited, with little gain achieved by U-64QAM over U-16QAM, and a large gap with respect to the SKR peak achievable by GG02; on the other hand, PAS provides a significant performance improvement and allows to better exploit alphabets with higher cardinality. PAS-16QAM largely outperforms both U-16QAM and U-64QAM, and PAS-64QAM provides a further improvement, closing the gap with GG02.

It is particularly interesting to explore how the maximum SKR varies with the transmission distance and verify if the relative behavior of the various protocols changes. For each distance, we define $\mathrm{SKR}_{\mathit{max}}$ as the peak of the SKR curve over the launch power, and introduce the dimensionless parameter $\mathrm{R}$:
\begin{equation}
    \mathrm{R} = \dfrac{\mathrm{SKR}_{\mathrm{max}}}{\mathrm{SKR}^{\mathrm{GM}}_{\mathrm{max}}}
\end{equation}
defined as the ratio between the $\mathrm{SKR}_{\mathrm{max}}$ value obtained for a given protocol, and that obtained for the GG02 protocol. This provides a measure of efficiency of a given protocol compared to GG02.

This efficiency parameter is reported in Fig. (\ref{fig:100km}) on the right. Compared to uniform signaling, PAS significantly improves the efficiency of the protocol and mitigates the decay of $\mathrm{R}$ at long distances, where channel loss dominates. The improvement is more pronounced for QAM constellations with higher cardinality due to the larger flexibility in shaping the input distribution. For example, PAS-64QAM reaches $95\%$ of the GG02 performance beyond $\unit[40]{km}$, while PAS-16QAM is already sufficient to reach almost $70\%$ at the same distance, outperforming both U-16QAM and U-64QAM, whose efficiency never exceeds $50\%$. In summary, PAS not only improves the SKR of QAM-based protocols, but also allows high-cardinality constellations like 64QAM to closely approach the performance of GG02, making practical protocols based on discrete modulations  more competitive in terms of maximum achievable SKR.

We now turn our attention to the role of excess noise, which --- along with losses --- affects performance in realistic scenarios. We fix the cardinality of the constellation (64QAM) and study the evolution of $\mathrm{SKR}_{\mathrm{max}}$ with distance for increasing excess noise levels $\xi\in\{0.01,0.03,0.05\}$ (in SNU).

\begin{figure*}[htp]
    \centering
    \includegraphics[width=1\textwidth]{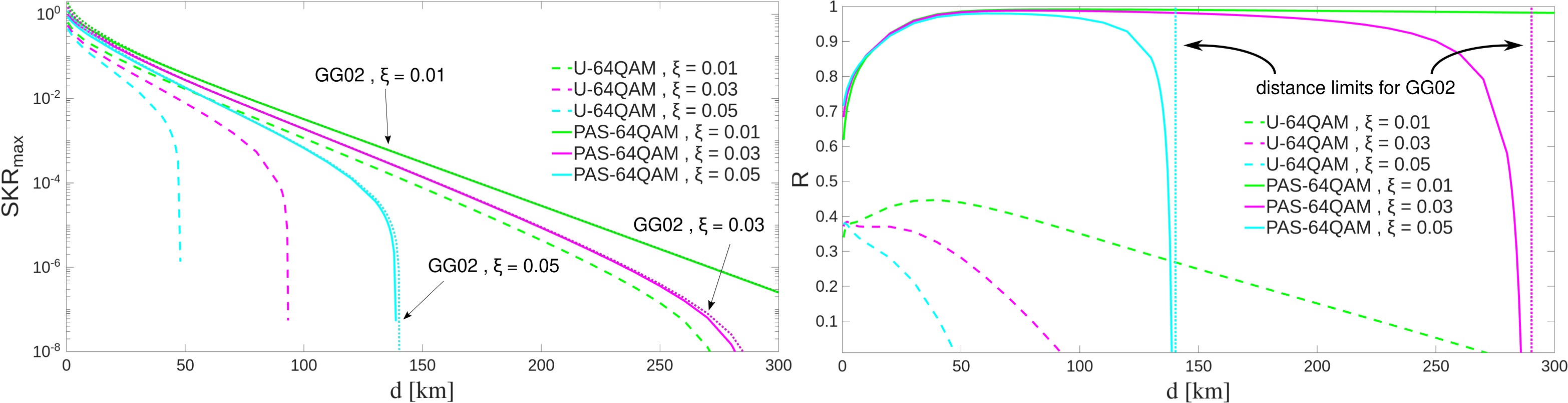}
    \caption{(Left) Maxima of SKR for the 64QAM with uniform and non-uniform input signaling in the range $\unit[0.5\text{--}300]{km}$ for different excess noise values $\xi$. GG02 curves are shown as dotted lines and are hardly distinguishable, as they are almost completely superimposed on the corresponding PAS curves. (Right) Corresponding values of $\mathrm{R}$ in the same range of distances and excess noise values; the dotted lines show the maximum achievable distances of GG02 for $\xi = 0.03$ (magenta) and $\xi = 0.05$ (light blue).}
    \label{fig:R_16_64}
\end{figure*}

As shown in Fig. (\ref{fig:R_16_64}) (left), the $\mathrm{SKR}_{\mathrm{max}}$ decays more rapidly with distance as $\xi$ increases, dropping to zero at progressively shorter distances. However, PAS still provides an effective solution, significantly slowing down the decay rate and extending the maximum distance compared to uniform signaling.

This resilience is also reflected in the behavior of the relative efficiency  $\mathrm{R}$, shown in Fig. (\ref{fig:R_16_64}) (right). At short distances, $\mathrm{R}$ is constrained by the constellation cardinality $M$, which limits the entropy of the source to $\log_{2}M$. Interestingly, in this regime, an increase of the excess noise level $\xi$ reduces the gap between discrete modulations and GG02. This is reflected in a moderate increase of $\mathrm{R}$ with $\xi$ for uniform signaling, and a slightly more pronounced increase for PAS. 

In general, PAS provides a substantial gain over uniform signaling at all noise levels. While shaped signaling cannot fully bridge the gap with GG02 at short distances --- due to the aforementioned entropy limitation --- it performs increasingly well in the medium-distance regime, where $\mathrm{R}$ approaches 1. In this regime, excess noise generally reduces the efficiency of discrete modulations relative to GG02. However, the advantage of PAS over uniform signaling becomes increasingly pronounced as the excess noise level rises, with the efficiency parameter $\mathrm{R}$ remaining close to 1 across a broad range of distances and noise levels. In fact, PAS-64QAM achieves almost the same distance limit as GG02, with its efficiency relative to GG02 only beginning to decline a few kilometers before that limit.

\begin{figure*}[htp]
    \centering\includegraphics[width=1\textwidth]{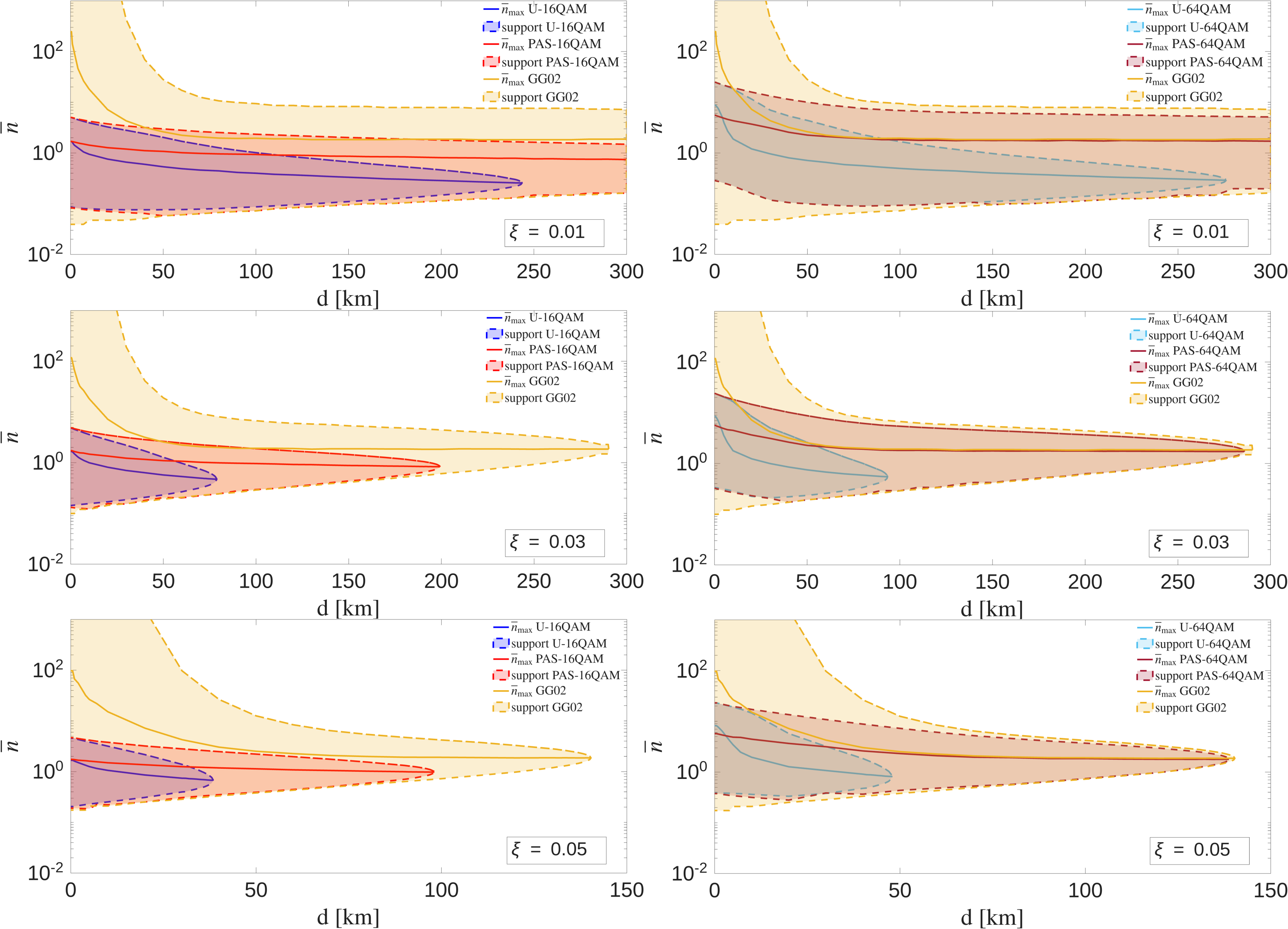}
    \caption{$\bar{n}_{\rm max}$ and mean number photons support of SKR curves at Alice's side for 16QAM (left) and for 64QAM (right) with both uniform and non-uniform signaling and with respect to the GG02 case at different excess noise values (from top to bottom $\xi=0.01,0.03,0.05$).}
    \label{fig:Contorni}
\end{figure*}

\subsection{Launch power}

\noindent A key design parameter in QKD systems is the launch power---typically expressed as the average number of photons per symbol $\bar{n}$ on Alice's side---which must be optimized to ensure secure operation at a given distance and noise level. Fig. (\ref{fig:Contorni}) reports, as a function of distance, the optimal launch power $\bar{n}_{\rm max}$ (solid lines) that maximizes the SKR, along with the operating power range (shaded areas), defined by the minimum and maximum $\bar{n}$ values (dotted lines) for which the SKR remains positive. The left and right panels correspond to 16QAM and 64QAM formats, respectively, with the GG02 reference shown in all plots. The top, middle, and bottom panels represent different levels of excess noise. For each distance, the corresponding power values on Bob's side is equal to $T (\bar{n} + \xi/2)$.

The GG02 protocol starts, at short distances (where channel losses are negligible), with a high optimal launch power (exceeding 100 photons per symbol) and a wide operating power range. As the distance increases, the optimal launch power initially decreases rapidly and then stabilizes at an almost constant value. At the same time, the power operating range gradually narrows around the optimal power, and eventually collapses when the protocol reaches its maximum transmission distance, which depends on the excess noise level. Discrete modulations exhibit a qualitatively similar trend, but with some relevant differences. First, the optimal launch power starts from much lower values at zero distance due to the entropy limitation of finite-size constellations. Additionally, both the power operating range and maximum achievable distance are generally reduced compared to GG02. These limitations are particularly evident for uniform signaling, with minimal improvement when moving from 16QAM to 64QAM. By contrast, the higher efficiency of PAS is reflected in both a higher optimal launch power and a broader operating range relative to uniform signaling. Notably, except for the first $\unit[70]{km}$ --- where the entropy constraint of the constellation still limits the performance --- PAS-64QAM closely follows the behavior of GG02, with practically the same optimal launch power, operating power range, and maximum distance.

\subsection{Maximum tolerable excess noise}

\begin{figure}[htp]
    \centering
    \includegraphics[width=1\columnwidth]{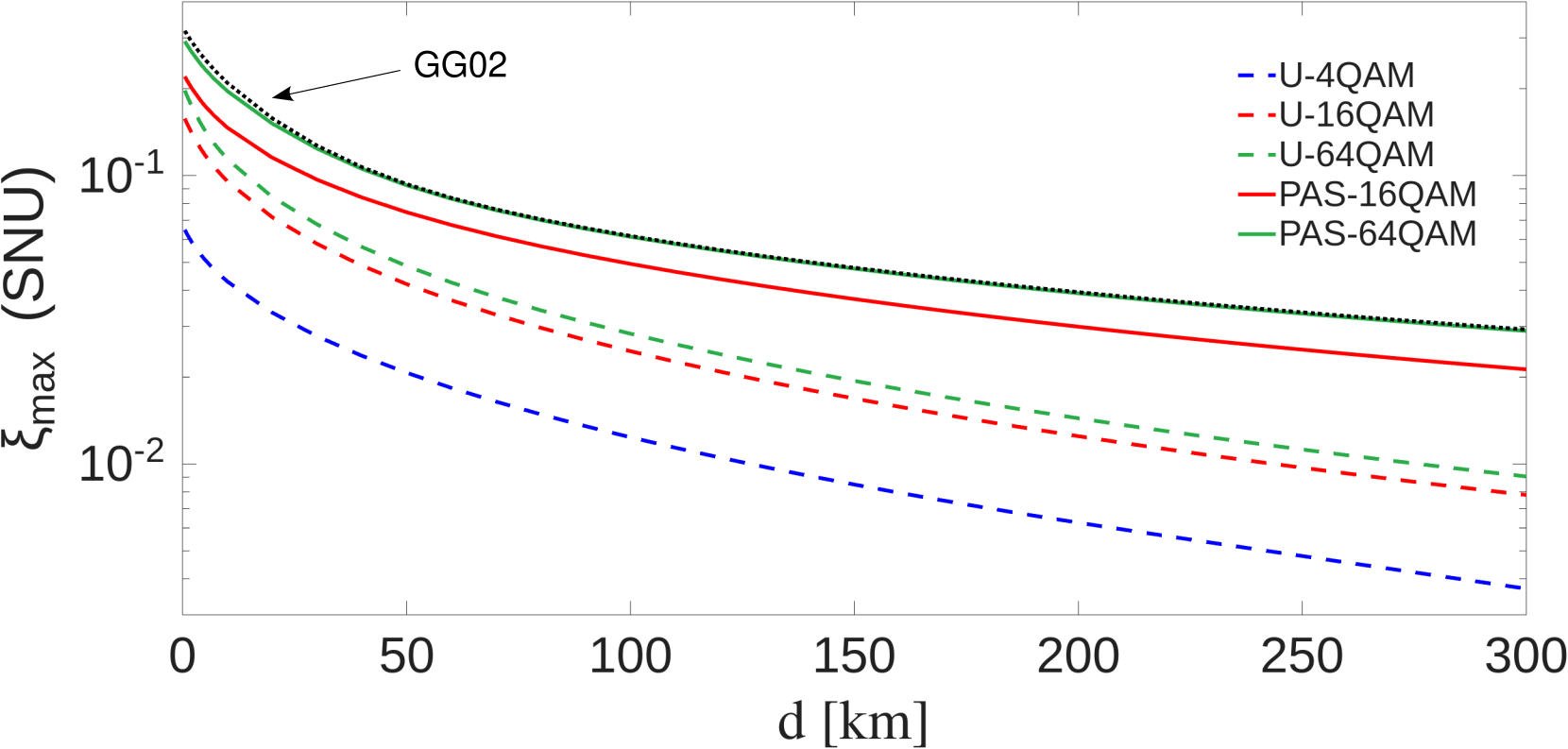}\caption{Tolerable excess noise $\xi_{\mathit{max}}$ for collective attacks, homodyne detection and reverse reconciliation in the range $\unit[0.5\text{--}300]{km}$ for 4QAM, 16QAM, 64QAM with uniform and non uniform signaling and for the GG02 protocol.}
    \label{fig:tolerable}
\end{figure}

\noindent We now assess the noise resilience of the proposed protocol by analyzing the maximum excess noise that can be tolerated as a function of the transmission distance. We define the maximum tolerable excess noise as the highest value of $\xi$ for which the SKR remains positive. Fig. (\ref{fig:tolerable}) compares this quantity for the GG02 protocol and for QAM-based protocols with uniform or shaped input distributions.

Consistent with the trends observed in previous sections, we find that the maximum tolerable excess noise increases with the cardinality $M$ of the constellation and decreases with transmission distance. This behavior reflects the opposing trends of the two terms in the SKR expression (\ref{SKR}): while the HI term  increases with noise, the MI term decreases with both noise and distance. As expected, uniform QAM is substantially less resilient to excess noise than GG02. However, PAS significantly improves the robustness of QAM-based protocols. Notably, at long distances, PAS-16QAM more than doubles the noise tolerance compared to both uniform 16QAM and 64QAM. Moreover, PAS-64QAM can tolerate nearly the same level of excess noise as GG02 at all distances beyond $\unit[40]{km}$.

\subsection{Wiretap channel security vs unconditional security}

An efficient evaluation of a generic protocol must take into account the level of security it guarantees, the performance it provides, and the feasibility of its implementation. Regarding the latter two, we have discussed above how our protocol is easily compatible with the classical telecommunication techniques currently exploited and what achievable rates, distances and noise resilience we can expect from it. Its security, instead, was untangled in the estimation of the HI term, where we assumed Eve is able to purify the system introducing additional modes that she can control.

In the following, we compare the results we have found here with those we obtained in our previous work \cite{notarnicola2023probabilistic}, in which we dealt with a QAM-based CV-QKD scheme, but under the different scenario of a wiretap quantum channel, where Eve is limited to beam-splitting attack. Although the wiretap channel represents a much more restrictive eavesdropping, a comparison between these two channel models can shed light on the role of the probabilistic shaping technique in a generic CV-QKD system with QAM modulation of the input symbols. It is therefore interesting to understand how the wiretap ($\mathit{w}$) and the linear quantum channel ($\ell$) hypotheses, characterized by different security levels, influence the performance of the same CV-QKD protocol and how the probabilistic shaping technique behaves as a resource in these different contexts. In particular, since we are interested in the advantages in the rate of communication (SKR), we explore these features only in a noiseless framework.

In a lossy (noiseless) wiretap channel scenario, Eve cannot alter the nature of the quantum signals: in fact, she can only collect the fraction of signals that has been lost during propagation; as a consequence, the channel in Eve's hands can be pictured as a beam splitter, whose output ports deliver the fraction $T$ of signals to Bob and the stolen fraction $1-T$ to Eve \cite{sych2010coherent}, \cite{pan2020secret}. In the linear noiseless quantum channel framework, the shot-noise is still the only resource that can be exploited by Eve to hide her interaction with the system, but Eve's actions are only constrained to preserve Alice and Bob’s statistics without being limited to Gaussian attacks, so that the attacker has more freedom in choosing the dilation map that allows her to maximize the knowledge on the system.

\begin{figure*}[htp]
    \centering
    \includegraphics[width=1\textwidth]{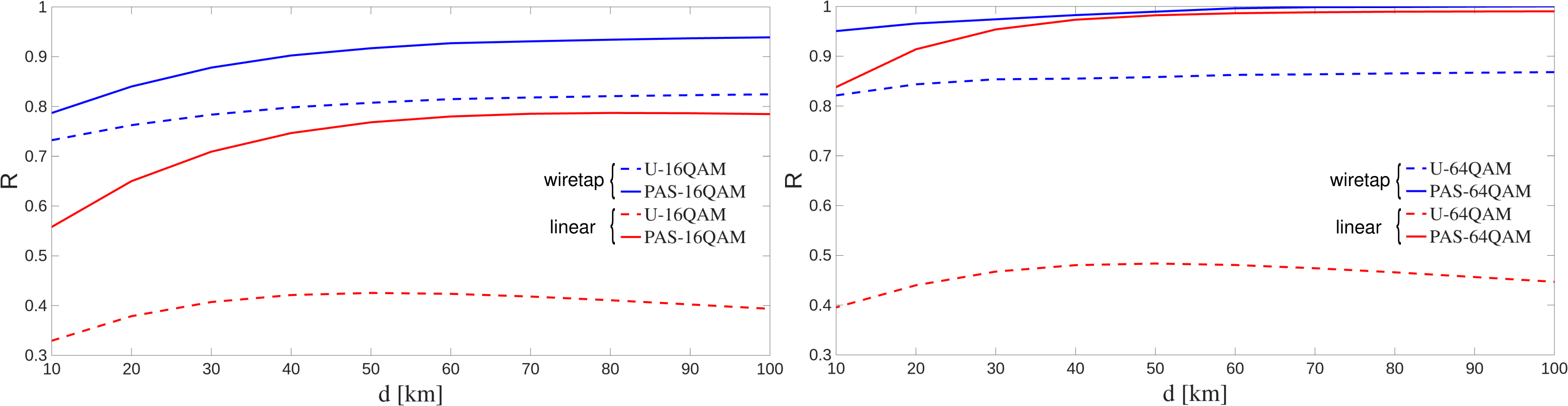}
    \caption{The parameter $\mathrm{R}$ for collective attacks, homodyne detection and reverse reconciliation in the range $\unit[10\text{--}100]{km}$ for 16QAM (left) and 64QAM (right) for a wiretap quantum channel and a linear quantum channel with no excess noise.}
    \label{fig:R_comparison}
\end{figure*}

As already seen, a good figure of merit for the evaluation of the achievable rate compared to that of the GG02 protocol is the parameter $\mathrm{R}$; in Fig. (\ref{fig:R_comparison}) we compared this quantity obtained for each model and for both 16QAM (left side) and 64QAM (right side). From a general point of view, the linear quantum framework ($\ell$) provides worse results than the wiretap one in terms of maximum rate. We can, in fact, notice how the curves related to the uniform signaling in ($\mathit{w}$) are very close to those obtained from a MB shaping of the input probabilities in ($\ell$): for example, U-16QAM ($\mathit{w}$) (blue dotted line) is able to outperform the PAS-16QAM obtained in ($\ell$) (red solid line). However, this feature does not occur in the 64QAM case: unlike before, here a larger statistical set contributes to a better effect of the PAS technique on the achievable rate of the key distribution.

Therefore, we can say that having a higher level of security against Eve's strategies and attacks in a linear quantum channel hypothesis has, in general, the price of lower performance. However, even if the values of $\mathrm{R}$ in ($\ell$) are noticeably lower than those in ($\mathit{w}$) for uniform signaling, shaping is still able to increase the system performance in the long distance regime, reducing the gap with GG02. It is important to note that the percentage increase in the rate is, proportionately, even larger for ($\ell$) than for ($\mathit{w}$); in particular, it can be easily seen that the performance of the PAS-64QAM ($\ell$) is almost equal to that of the PAS-64QAM ($\mathit{w}$) after a distance of $\unit[40]{km}$.

\section{Conclusions}

\noindent In this work, we have investigated a QAM-based CV-QKD system with uniform and non-uniform signaling (PAS technique using a Maxwell--Boltzmann distribution) within the framework of homodyne detection and the asymptotic regime of infinite key length. We focused on collective attacks, which are the most powerful class of attacks in the asymptotic limit, and adopted the reverse reconciliation scheme since it enables longer communication distances between the legitimate parties. In particular, we studied the system under a linear quantum channel model---which provides a simple and effective description of the propagation of quantum signals in an optical fiber link---and analyzed its behavior under the purification attack, which represents the most powerful strategy available to Eve to extract information from the system. Using the GG02 protocol as a benchmark, we evaluated the system performance in terms of achievable secure key rates and maximum transmission distances for different levels of excess noise, and assessed its robustness against noise.

We found that the combination of QAM modulation and PAS based on the Maxwell-Boltzmann distribution consistently outperforms the corresponding uniform distribution protocol for all the distances and excess noise values considered in this work. In particular, PAS-64QAM proved to be an effective and practical solution to closely approach the performance of the GG02 protocol, especially in the long-distance regime. Indeed, beyond approximately $\unit[40]{km}$, PAS-64QAM achieves nearly the same secure key rates and exhibits the same excess noise tolerance compared to the ideal GG02 protocol. Furthermore, PAS significantly enhances the performance of QAM-based schemes at any distance, with the simpler PAS-16QAM outperforming the more complex uniform 64QAM and showing increased robustness to excess noise.

Moreover, we analyzed the impact of PAS under different assumptions on Eve’s capabilities. To this end, we have compared the results obtained for a noiseless linear quantum channel model with those corresponding to a lossy wiretap channel scenario, in which Eve is limited to accessing only the fraction of losses introduced by imperfections in the optical fiber link. In general, we observed that the use of discrete modulation assisted by a uniform probability of input symbols leads to lower achievable key rates in a noiseless linear quantum channel compared to the lossy wiretap model. Nevertheless, within the noiseless linear quantum channel framework, PAS not only provides an effective means to mitigate this performance degradation, but also yields a higher percentage gain in terms of secure key rates compared to the lossy wiretap model. These features are particularly evident from the observed good agreement between the maximum secure key rates obtained under the two channel models for PAS-64QAM in the long-distance regime.

It should be emphasized that, for non-Gaussian protocols such as the one considered here, the optimal attack strategy for Eve remains an open problem, and this paves the way for future investigation of the role of non-Gaussian attacks in the security analysis of DM CV-QKD protocols. However, if on one side the linear channel assumption provides Eve with a less restricted operational framework compared to the lossy wiretap model, on the other side it still constitutes a limitation on her capabilities. To further strengthen the security assessment of discrete-modulation CV-QKD, future work will address the most general scenario of a non-linear quantum channel, characterized by a non-linear input-output relation between the quadrature operators.

\section*{Acknowledgments}

This work was supported in part by the by PNRR MUR project PE0000023-NQSTI, and by the National Operational Programme on Research and Innovation 2014–2020 - FSE REACT EU “Azione IV.5 Dottorati su tematiche Green”.

\bibliographystyle{IEEEtran}
\bibliography{sample}

@article{weedbrook2012gaussian,
  title={Gaussian quantum information},
  author={Weedbrook, Christian and Pirandola, Stefano and Garc{\'\i}a-Patr{\'o}n, Ra{\'u}l and Cerf, Nicolas J and Ralph, Timothy C and Shapiro, Jeffrey H and Lloyd, Seth},
  journal={Reviews of Modern Physics},
  volume={84},
  number={2},
  pages={621--669},
  year={2012},
  publisher={APS}
}

@inproceedings{maurer1999information,
  title={Information-theoretic cryptography},
  author={Maurer, Ueli},
  booktitle={Annual International Cryptology Conference},
  pages={47--65},
  year={1999},
  organization={Springer}
}

@article{duvsek2006quantum,
  title={Quantum cryptography},
  author={Du{\v{s}}ek, Miloslav and L{\"u}tkenhaus, Norbert and Hendrych, Martin},
  journal={Progress in optics},
  volume={49},
  pages={381--454},
  year={2006},
  publisher={Elsevier New York}
}

@article{gisin2002quantum,
  title={Quantum cryptography},
  author={Gisin, Nicolas and Ribordy, Gr{\'e}goire and Tittel, Wolfgang and Zbinden, Hugo},
  journal={Reviews of modern physics},
  volume={74},
  number={1},
  pages={145},
  year={2002},
  publisher={APS}
}

@article{scarani2009security,
  title={The security of practical quantum key distribution},
  author={Scarani, Valerio and Bechmann-Pasquinucci, Helle and Cerf, Nicolas J and Du{\v{s}}ek, Miloslav and L{\"u}tkenhaus, Norbert and Peev, Momtchil},
  journal={Reviews of modern physics},
  volume={81},
  number={3},
  pages={1301},
  year={2009},
  publisher={APS}
}

@article{martin2017introduction,
  title={Introduction to quantum key distribution},
  author={Martin, Vicente and Martinez-Mateo, Jesus and Peev, Momtchil},
  journal={Wiley Encyclopedia of Electrical and Electronics Engineering},
  volume={1},
  pages={1--17},
  year={2017}
}

@article{pirandola2020advances,
  title={Advances in quantum cryptography},
  author={Pirandola, Stefano and Andersen, Ulrik L and Banchi, Leonardo and Berta, Mario and Bunandar, Darius and Colbeck, Roger and Englund, Dirk and Gehring, Tobias and Lupo, Cosmo and Ottaviani, Carlo and others},
  journal={Advances in optics and photonics},
  volume={12},
  number={4},
  pages={1012--1236},
  year={2020},
  publisher={Optica Publishing Group}
}

@book{wolf2021quantum,
  title={Quantum key distribution},
  author={Wolf, Ramona},
  year={2021},
  publisher={Springer}
}

@article{zhang2024continuous,
  title={Continuous-variable quantum key distribution system: Past, present, and future},
  author={Zhang, Yichen and Bian, Yiming and Li, Zhengyu and Yu, Song and Guo, Hong},
  journal={Applied Physics Reviews},
  volume={11},
  number={1},
  year={2024},
  publisher={AIP Publishing}
}

@article{kaur2021asymptotic,
  title={Asymptotic security of discrete-modulation protocols for continuous-variable quantum key distribution},
  author={Kaur, Eneet and Guha, Saikat and Wilde, Mark M},
  journal={Physical Review A},
  volume={103},
  number={1},
  pages={012412},
  year={2021},
  publisher={APS}
}

@article{laudenbach2018continuous,
  title={Continuous-variable quantum key distribution with Gaussian modulation—the theory of practical implementations},
  author={Laudenbach, Fabian and Pacher, Christoph and Fung, Chi-Hang Fred and Poppe, Andreas and Peev, Momtchil and Schrenk, Bernhard and Hentschel, Michael and Walther, Philip and H{\"u}bel, Hannes},
  journal={Advanced Quantum Technologies},
  volume={1},
  number={1},
  pages={1800011},
  year={2018},
  publisher={Wiley Online Library}
}

@article{abushgra2022variations,
  title={Variations of QKD protocols based on conventional system measurements: A literature review},
  author={Abushgra, Abdulbast A},
  journal={Cryptography},
  volume={6},
  number={1},
  pages={12},
  year={2022},
  publisher={MDPI}
}

@article{grosshans2002continuous,
  title={Continuous variable quantum cryptography using coherent states},
  author={Grosshans, Fr{\'e}d{\'e}ric and Grangier, Philippe},
  journal={Physical review letters},
  volume={88},
  number={5},
  pages={057902},
  year={2002},
  publisher={APS}
}

@book{cerf2007quantum,
  title={Quantum information with continuous variables of atoms and light},
  author={Cerf, Nicolas J and Leuchs, Gerd and Polzik, Eugene S},
  year={2007},
  publisher={World Scientific}
}

@article{jouguet2012analysis,
  title={Analysis of imperfections in practical continuous-variable quantum key distribution},
  author={Jouguet, Paul and Kunz-Jacques, S{\'e}bastien and Diamanti, Eleni and Leverrier, Anthony},
  journal={Physical Review A—Atomic, Molecular, and Optical Physics},
  volume={86},
  number={3},
  pages={032309},
  year={2012},
  publisher={APS}
}

@article{leverrier2009unconditional,
  title={Unconditional security proof of long-distance continuous-variable quantum key distribution with discrete modulation},
  author={Leverrier, Anthony and Grangier, Philippe},
  journal={Physical review letters},
  volume={102},
  number={18},
  pages={180504},
  year={2009},
  publisher={APS}
}

@article{djordjevic2019discretized,
  title={On the discretized Gaussian modulation (DGM)-based continuous variable-QKD},
  author={Djordjevic, Ivan B},
  journal={IEEE Access},
  volume={7},
  pages={65342--65346},
  year={2019},
  publisher={IEEE}
}

@article{fehenberger2018information,
  title={Information rates of probabilistically shaped coded modulation for a multi-span fiber-optic communication system with 64QAM},
  author={Fehenberger, Tobias},
  journal={Optics Communications},
  volume={409},
  pages={2--6},
  year={2018},
  publisher={Elsevier}
}

@article{kschischang1993optimal,
  title={Optimal nonuniform signaling for Gaussian channels},
  author={Kschischang, Frank R and Pasupathy, Subbarayan},
  journal={IEEE Transactions on Information Theory},
  volume={39},
  number={3},
  pages={913--929},
  year={1993},
  publisher={IEEE}
}

@article{fehenberger2016probabilistic,
  title={On probabilistic shaping of quadrature amplitude modulation for the nonlinear fiber channel},
  author={Fehenberger, Tobias and Alvarado, Alex and B{\"o}cherer, Georg and Hanik, Norbert},
  journal={Journal of Lightwave Technology},
  volume={34},
  number={21},
  pages={5063--5073},
  year={2016},
  publisher={IEEE}
}

@article{forney1984efficient,
  title={Efficient modulation for band-limited channels},
  author={Forney, GD and Gallager, R and Lang, G and Longstaff, F and Qureshi, S},
  journal={IEEE journal on selected areas in communications},
  volume={2},
  number={5},
  pages={632--647},
  year={1984},
  publisher={IEEE}
}

@article{bocherer2015bandwidth,
  title={Bandwidth efficient and rate-matched low-density parity-check coded modulation},
  author={B{\"o}cherer, Georg and Steiner, Fabian and Schulte, Patrick},
  journal={IEEE Transactions on communications},
  volume={63},
  number={12},
  pages={4651--4665},
  year={2015},
  publisher={IEEE}
}

@article{notarnicola2023probabilistic,
  title={Probabilistic amplitude shaping for continuous-variable quantum key distribution with discrete modulation over a wiretap channel},
  author={Notarnicola, Michele N and Olivares, Stefano and Forestieri, Enrico and Parente, Emanuele and Pot{\`\i}, Luca and Secondini, Marco},
  journal={IEEE Transactions on Communications},
  year={2023},
  publisher={IEEE}
}

@article{ghorai2019asymptotic,
  title={Asymptotic security of continuous-variable quantum key distribution with a discrete modulation},
  author={Ghorai, Shouvik and Grangier, Philippe and Diamanti, Eleni and Leverrier, Anthony},
  journal={Physical Review X},
  volume={9},
  number={2},
  pages={021059},
  year={2019},
  publisher={APS}
}

@book{cover1999elements,
  title={Elements of information theory},
  author={Cover, Thomas M},
  year={1999},
  publisher={John Wiley \& Sons}
}

@article{forney1989multidimensional,
  title={Multidimensional constellations. I. Introduction, figures of merit, and generalized cross constellations},
  author={Forney, G David and Wei, L-F},
  journal={IEEE journal on selected areas in communications},
  volume={7},
  number={6},
  pages={877--892},
  year={1989},
  publisher={IEEE}
}

@article{cariolaro2015quantum,
  title={Quantum Communications Systems},
  author={Cariolaro, Gianfranco and Cariolaro, Gianfranco},
  journal={Quantum Communications},
  pages={281--359},
  year={2015},
  publisher={Springer}
}

@incollection{secondini2020information,
  title={Information capacity of optical channels},
  author={Secondini, Marco},
  booktitle={Optical Fiber Telecommunications VII},
  pages={867--920},
  year={2020},
  publisher={Elsevier}
}

@article{lin2019asymptotic,
  title={Asymptotic security analysis of discrete-modulated continuous-variable quantum key distribution},
  author={Lin, Jie and Upadhyaya, Twesh and L{\"u}tkenhaus, Norbert},
  journal={Physical Review X},
  volume={9},
  number={4},
  pages={041064},
  year={2019},
  publisher={APS}
}

@article{denys2021explicit,
  title={Explicit asymptotic secret key rate of continuous-variable quantum key distribution with an arbitrary modulation},
  author={Denys, Aur{\'e}lie and Brown, Peter and Leverrier, Anthony},
  journal={Quantum},
  volume={5},
  pages={540},
  year={2021},
  publisher={Verein zur F{\"o}rderung des Open Access Publizierens in den Quantenwissenschaften}
}

@article{garcia2006unconditional,
  title={Unconditional Optimality of Gaussian Attacks against Continuous-Variable Quantum Key Distribution},
  author={Garc{\'\i}a-Patr{\'o}n, Ra{\'u}l and Cerf, Nicolas J},
  journal={Physical review letters},
  volume={97},
  number={19},
  pages={190503},
  year={2006},
  publisher={APS}
}

@article{braunstein2005quantum,
  title={Quantum information with continuous variables},
  author={Braunstein, Samuel L and Van Loock, Peter},
  journal={Reviews of modern physics},
  volume={77},
  number={2},
  pages={513--577},
  year={2005},
  publisher={APS}
}

@article{navascues2006optimality,
  title={Optimality of Gaussian attacks in continuous-variable quantum cryptography},
  author={Navascu{\'e}s, Miguel and Grosshans, Fr{\'e}d{\'e}ric and Acin, Antonio},
  journal={Physical review letters},
  volume={97},
  number={19},
  pages={190502},
  year={2006},
  publisher={APS}
}

@article{lupo2022quantum,
  title={Quantum key distribution with nonideal heterodyne detection: composable security of discrete-modulation continuous-variable protocols},
  author={Lupo, Cosmo and Ouyang, Yingkai},
  journal={PRX Quantum},
  volume={3},
  number={1},
  pages={010341},
  year={2022},
  publisher={APS}
}

@article{diamanti2015distributing,
  title={Distributing secret keys with quantum continuous variables: principle, security and implementations},
  author={Diamanti, Eleni and Leverrier, Anthony},
  journal={Entropy},
  volume={17},
  number={9},
  pages={6072--6092},
  year={2015},
  publisher={MDPI}
}

@article{olivares2012quantum,
  title={Quantum optics in the phase space: A tutorial on Gaussian states},
  author={Olivares, Stefano},
  journal={The European Physical Journal Special Topics},
  volume={203},
  number={1},
  pages={3--24},
  year={2012},
  publisher={Springer}
}

@article{sych2010coherent,
  title={Coherent state quantum key distribution with multi letter phase-shift keying},
  author={Sych, Denis and Leuchs, Gerd},
  journal={New Journal of Physics},
  volume={12},
  number={5},
  pages={053019},
  year={2010},
  publisher={IOP Publishing}
}

@article{pan2020secret,
  title={Secret-key distillation across a quantum wiretap channel under restricted eavesdropping},
  author={Pan, Ziwen and Seshadreesan, Kaushik P and Clark, William and Adcock, Mark R and Djordjevic, Ivan B and Shapiro, Jeffrey H and Guha, Saikat},
  journal={Physical Review Applied},
  volume={14},
  number={2},
  pages={024044},
  year={2020},
  publisher={APS}
}

@article{proakis2001digital,
  title={Digital Communications 4th ed. McGraw-Hill},
  author={Proakis, J and Salehi, M},
  journal={New York},
  year={2001}
}

@article{winzer2012high,
  title={High-spectral-efficiency optical modulation formats},
  author={Winzer, Peter J},
  journal={Journal of lightwave technology},
  volume={30},
  number={24},
  pages={3824--3835},
  year={2012},
  publisher={OSA}
}

@article{buchali2015rate,
  title={Rate adaptation and reach increase by probabilistically shaped 64-QAM: An experimental demonstration},
  author={Buchali, Fred and Steiner, Fabian and B{\"o}cherer, Georg and Schmalen, Laurent and Schulte, Patrick and Idler, Wilfried},
  journal={Journal of lightwave technology},
  volume={34},
  number={7},
  pages={1599--1609},
  year={2015},
  publisher={IEEE}
}

@inproceedings{origlia2025soft,
  title={Soft Reverse Reconciliation for Discrete Modulations},
  author={Origlia, Marco and Secondini, Marco},
  booktitle={2025 14th International ITG Conference on Systems, Communications and Coding (SCC)},
  pages={1--6},
  year={2025},
  organization={IEEE}
}

@article{ferraro2005gaussian,
  title={Gaussian states in continuous variable quantum information},
  author={Ferraro, Alessandro and Olivares, Stefano and Paris, Matteo GA},
  journal={arXiv preprint quant-ph/0503237},
  year={2005}
}

@article{holevo2012quantum,
  title={Quantum channels and their entropic characteristics},
  author={Holevo, Alexander S and Giovannetti, Vittorio},
  journal={Reports on progress in physics},
  volume={75},
  number={4},
  pages={046001},
  year={2012},
  publisher={IOP Publishing}
}

@article{notarnicola2024quantum,
  title={Quantum communications in continuous variable systems},
  author={Notarnicola, Michele N and others},
  year={2024},
  publisher={Universit{\`a} degli Studi di Milano}
}

@article{notarnicola2023long,
  title={Long-distance continuous-variable quantum key distribution with feasible physical noiseless linear amplifiers},
  author={Notarnicola, Michele N and Olivares, Stefano},
  journal={Physical Review A},
  volume={108},
  number={2},
  pages={022404},
  year={2023},
  publisher={APS}
}

@phdthesis{leverrier2009theoretical,
  title={Theoretical study of continuous-variable quantum key distribution},
  author={Leverrier, Anthony},
  year={2009},
  school={T{\'e}l{\'e}com ParisTech}
}

\end{document}